# Network Pharmacology Reveals HSPA1A/BST2 as Potential Targets of Ci Bai Capsule's Active Compounds Intervening in Leukopenia

Dingfan Zhang[#, a], Congshu Huang[#, d], Lei Zhou[#, b], Boyang Wang[a], Wei Zhou[b], Tiantian Xia[e], Pan Shen[*, b], Shao Li[*, a] & Yue Gao[*, bc]

[a]Institute for TCM-X, MOE Key Laboratory of Bioinformatics/Bioinformatics Division, BNRIST, Department of Automation, Tsinghua University, Beijing 100084, China

[b]Department of Pharmaceutical Sciences, Beijing Institute of Radiation Medicine, Beijing 100850, China

[c]State Key Laboratory of Kidney Diseases, Chinese PLA General Hospital, Beijing 100853, China

[d]College of Traditional Chinese Medicine, Henan University of Chinese Medicine, Zhengzhou 455046, China

[e]Medical College, Qinghai University, Xining 810016, China

* Correspondence should be addressed to Pan Shen (spluto@foxmail.com), Shao Li (shaoli@mail.tsinghua.edu.cn) and Yue Gao (gaoyue@bmi.ac.cn)

# These authors should be regarded as joint First Authors

**Abstract**

Background: Radiation-induced leukopenia caused by low-dose exposure is frequently associated with Traditional Chinese Medicine (TCM) syndromes like "blood deficiency" and "fatigue syndrome". Ci Bai Capsule (CB) has been reported to enhance white blood cell levels; however, its mechanisms and bioactive compounds remain unclear.

Aim: This study aimed to identify the bioactive compounds group of CB and elucidate its potential mechanisms in radiation-induced leukopenia.

Methods: Syndrome-related data were gathered from SYMMAP and CTD database. CB's target profile is predicted by DrugCIPHER. Network pharmacology approaches were employed to identify active compounds and related pathways. Experimental validation was conducted through flow cytometry and RNA-sequencing in both ex vivo and in vivo models.

Results: A total of 22 pathways related to cellular processes, immune responses, and signal transduction were identified. Five key bioactive compounds (kaempferol-3-glucorhamnoside, syringin, schisandrin, 3-hydroxytyrosol 3-O-glucoside and salidroside) were found to significantly modulate syndrome-related pathways. Optimal dosing of this compound combination enhanced leukocyte counts and splenic immune cell proliferation in irradiated mice. Transcriptomic analysis revealed that the compounds exert regulatory effects on PP1A, RB, CDK4/6, CDK2, and CDK1, thereby modulating downstream immune and hematopoietic markers such as MNDA, BST2, and HSPA1A.

Conclusion: Our findings suggest that CB mitigates radiation-induced leukopenia by enhancing immune and hematopoietic recovery, offering a promising therapeutic approach for managing radiation-related hematological disorders.

**Keywords:** Ci Bai Capsule; Network target; Leukopenia; Transcriptomics; RNA-seq analysis

**Abbreviations**

Ci Bai Capsule: CB;

Traditional Chinese Medicine: TCM;

White blood cell: WBC;

Scleromitrion diffusum: BHSSC;

Eleutherococcus senticosus: CWJ;

Poria: FL;

Atractylodes macrocephala Koidz: BZ;

Codonopsis radix: DS;

Reynoutria japonica: HZ;

Lycium chinense: GQZ;

Cuscuta: TSZ;

Schisandra chinensis: WWZ;

Ligustri lucidi fructus: NZZ;

Febrile Disease: FD;

Weakened Spleen and Kidneys: WSK;

Insufficient Blood: IB;

Chronic Deficiency or Fatigue Syndrome: CDFS;

Sovereign: JUN;

Minister: CHEN;

Assistant: ZUO/SHI.

## Introduction

Radiation-induced leukopenia, occurring in over 20% of radiotherapy patients (Venkatesulu et al., 2018), manifests as critically low white blood cell (WBC) counts due to radiation exposure (Nelson, 2016; Yang et al., 2023), and severely compromises immune function and patient outcomes (Cella et al., 2024; Chen et al., 2023; Pham et al., 2023). This disruption arises from dual radiation effects: directly hematopoietic stem cell damage and microenvironmental impairment in spleen and bone marrow, which governs WBC storage, proliferation, and activation (Newton et al., 2019; Velardi et al., 2021). Thus, developing protective measures is crucial. However, current commonly used therapeutic options remain limited in variety and demonstrate suboptimal efficacy and side effects (Bayne et al., 2012; Chen et al., 2023; Dai et al., 2010).

Ci Bai Capsule (CB), a traditional Chinese medicine (TCM) formula derived from the ancient prescriptions "Si Jun Zi Tang" and "Wu Zi Diffractive Zong Pill", emerges as a promising alternative treatment due to its immunomodulatory effects. Clinical evidence has confirmed the CB's efficacy in treating leukocyte reduction caused by radiation and chemotherapy (Hu et al., 2024). The core herbs in CB, including *Scleromitrion diffusum* (BHSSC), *Eleutherococcus senticosus* (CWJ) and *Poria* (FL), work synergistically to tonify Qi (Lee et al., 2019), nourish blood (Li et al., 2023), and enhance hematopoiesis (Provalova et al., 2002). Consequently, CB is deemed capable of strengthening the spleen, restoring immune balance, and boosting the body's natural defenses, thereby demonstrating particular effectiveness in managing leukopenia. However, the unclear pharmacological mechanisms hinder its clinical application and international recognition.

Network pharmacology offers a systematic and holistic approach to address the complexity of multicomponent herbal medicines by integrating molecular interaction data with pharmacological insights (Shao and Zhang, 2013; Zhang et al., 2024; Zhao et al., 2023; 李梢, 2011). The "Network Target" theory enables the identification of bioactive compounds and their corresponding therapeutic mechanisms, providing a robust framework for exploring the efficacy of traditional formulas (Li, 2015; 2021;

Zhang et al., 2024). This approach enables the resolution of the therapeutic potential of CB in treating radiation-induced leukopenia by combining network pharmacology with TCM syndrome analysis, aiming to uncover the active compound combinations and underlying mechanisms that contribute to its efficacy.

In this study, we systematically investigated the therapeutic mechanisms of CB against radiation-induced leukopenia through network pharmacology analysis. By analyzing CB's multi-component interactions, we identified potential regulatory networks within radiation-affected immune and hematopoietic systems. A pharmacological screening method incorporating TCM syndrome characteristics of leukopenia was applied to prioritize key bioactive compounds. Experimental validation further revealed CB's dual mechanisms of leukogenic action. These findings clarify CB's therapeutic basis while providing a methodological approach for evaluating complex herbal formulations, contributing to the development of mechanism-informed TCM therapies for radiation injuries.

## Methods and Materials

### Data collection

Data of syndromes were gathered from the SYMMAP (Wu et al., 2019) database, identifying OMIM standard entries related to TCM syndromes associated with leukopenia. For example, within the framework of TCM theory, the TCM syndrome "weakened function of the spleen and kidneys" in SYMMAP is related with phenotypes like Asthenia, Diarrhea, vision_low and Mental Deficiency. Each phenotype has its own standard Mesh term. Furthermore, based on the comparative toxicogenomics database (CTD) (Davis et al., 2021), genes linked to each Mesh term were identified and designated as genes associated with the TCM syndrome.

### Target prediction and validation

The DrugCIPHER algorithm (Zhao and Li, 2010), which is based on chemical similarity and network-driven drug target prediction, was employed to predict the genome-wide targets of each chemical constituent in CB. The top 100 predicted targets for each compound were considered as the compound's target profile. To explore the co-occurrence relationships between each chemical component and its predicted targets, abstracts from the PubMed database and drug-target data from the PubChem database were utilized. These co-occurrences were systematically organized to assess the extent of literature coverage for the predicted targets. The existing literature coverage was then leveraged to evaluate the predictive accuracy of each compound's target identification. Accuracy was calculated according to the following formula:

$$\text{Accuracy} = \left(\frac{Number\ of\ intersections\ between\ predicted\ targets\ and\ reported\ biomolecules}{Total\ number\ predicted\ targets}\right) \times 100\% \qquad (1)$$

**Network target analysis and multilayer network construction**

Statistical analyses were performed under the hypothesis that targets present in a higher number of major component compounds within formulations are more likely to represent key therapeutic targets. This approach facilitated the construction of a holistic target profile for the formula (Liang et al., 2014), allowing for the identification of targets that may play a pivotal role in its pharmacological effects. The results provided an integrated overview of the target landscape associated with the TCM formulation:

$$p_r(K = k) = \sum_{A \in F_k} \prod_{i \in A} p_i \prod_{j \in A^c} (1 - p_j) \qquad (2)$$

The identified holistic targets were subsequently subjected to KEGG pathway (Kanehisa, 2002) and gene ontology (GO) term (Ashburner et al., 2000) enrichment analyses to identify key biological pathways and processes potentially involved in the treatment of leukopenia by CB. Integrating these results with RNA-seq data, relevant pathways and biological processes associated with leukopenia were selected from the enriched targets of CB. These pathways and processes were classified into three distinct biological modules according to their functional roles.

To further elucidate the molecular interactions, a protein-protein interaction (PPI) network was constructed using the STRING database (Mering et al., 2003), focusing on proteins enriched in each module's pathways. The resulting network was visualized using Cytoscape (v3.9.1) software (Smoot et al., 2011). Based on TCM principles of compatibility, the herbal components of CB were categorized into three roles: sovereign (JUN), minister (CHEN), assistant (ZUO/SHI). The holistic targets associated with each category of herb were calculated and linked to the corresponding target columns. Furthermore, the association of these targets with specific pathways within the biological modules highlighted the enrichment of these targets in particular signaling pathways relevant to the treatment of leukopenia.

**Selection of potential bioactive compounds group (BCG)**

Compounds with significant therapeutic effects on the TCM syndromes of leukopenia were identified and selected to exemplify the therapeutic actions of CB. Based on previously collected "syndrome-genes" data, four distinct gene datasets corresponding to TCM syndromes were established, including febrile disease (FD), weakened spleen and kidneys (WSK), insufficient blood (IB), and chronic deficiency or fatigue syndrome (CDFS). The enrichment of the compounds' target profiles within each dataset was evaluated using hypergeometric distribution principles, as outlined in equation (3). A significant result ($P$ value < 0.05) indicated a potential therapeutic effect of the compound on the respective syndrome.

Furthermore, compounds that showed significant enrichment in any of the datasets were selected for further analysis based on their roles as sovereign ingredients and their inclusion as quality control compounds in the Chinese Pharmacopoeia. These criteria ensured the relevance of the selected compounds to the clinical application and quality standards of CB.

$$P = 1 - \sum_{i=0}^{x-1} \frac{\binom{M}{i}\binom{N-M}{K-i}}{\binom{N}{K}} \quad (3)$$

**Animal irradiation and treatment**

Eighty male C57BL/6 mice of 8 weeks weighing 20 ± 2 g were purchased from GemPharmatech Co. ltd (Nanjing, China). All animal experiments conducted in this study received approval from the Animal Care and Use Committee of the Academy's Animal Center. The mice were maintained under a controlled environment with a 12-hour light/dark cycle at room temperature (25 ± 2°C) and humidity (55 ± 5%), with unrestricted access to food and water. The $^{60}$Co gamma irradiation source used in the experiment was the Beijing Institute of Radiation Medicine. The mice were housed in a special case behind a shielded lead plate and irradiated with $^{60}$Co gamma irradiation. A single irradiation dose rate of 2.98 cGy/min, 4m away from the radiation source, irradiation dose of 0.1 Gy, at the same position for 5 consecutive days. The mice were divided into normal control group, irradiation group and irradiation administration group. The control group underwent no treatment, the irradiation administration group was given drug gavage immediately after the completion of irradiation for five consecutive days, and the irradiation group was given normal saline gavage.

**Uniform design of experimental scheme**

Using $U_6(6^5)$ uniform design table, we selected Syringin, Salidroside, Kaempferol, Schisandrol A, and Cimidahurinine as the examined factors. Each factor was assigned 6 levels, and the dosage ranges for Syringin, Salidroside, Kaempferol, Schisandrol A, and Cimidahurinine were determined based on literature to be 10-60 $mg \cdot kg^{-1}$、25-85 $mg \cdot kg^{-1}$、5-40 $mg \cdot kg^{-1}$、5-100 $mg \cdot kg^{-1}$、1-41 $mg \cdot kg^{-1}$, respectively. Using the uniform design table (Table 1), we designed a dosage combination experiment plan.

**Mouse blood cell level detection**

The mice were anesthetized with pentobarbital sodium, and after the left eye was removed and blood was collected, it was treated with anticoagulant and analyzed using an automatic hematocrit analyzer.

**Flow cytometric analysis**

On the sixth day after irradiation, the mice were euthanized. The thymus and spleen were collected separately and ground to collect cells. The collected cells were washed once and then re-suspended in 100 μL PBS containing 2% fetal bovine serum (FBS). For staining of lymphocytes in the thymus and spleen, the cells were incubated with biotin-conjugated antibodies against CD4 and CD8 for 30 minutes, followed by incubation with streptavidin-conjugated secondary antibodies and CD45 for 30 min after one wash. The cells were then washed once and re-suspended in PBS containing 0.1% bovine serum albumin. Five minutes before analyzing the cells on the Aria III (BD Biosciences), 5 μL of 7-AAD viability staining solution was used.

**RNA-seq analysis**

Total RNA was used for RNA sample preparation, with mRNA isolated via poly-T oligo-attached magnetic beads. Fragmentation was performed under elevated temperature with divalent cations in the First Strand Synthesis Reaction Buffer. First strand cDNA synthesis utilized random hexamer primers and M-MuLV Reverse Transcriptase, followed by second strand synthesis with DNA Polymerase I and RNase H. Overhangs were converted to blunt ends, and after adenylation of the 3' ends, adaptors with a hairpin loop structure were ligated. The cDNA fragments were size-selected to a length range of 370-420 bp using the AMPure XP system, followed by PCR amplification with Phusion High-Fidelity DNA polymerase and universal primers. PCR products were purified and library quantification was carried out using Qubit. The effective library concentration was confirmed by qRT-PCR to ensure quality. Qualified libraries were pooled and sequenced based on the required concentration and data output.

For bioinformatics analysis, raw sequencing data were processed using fastp (v0.23.1) to assess quality. The sequence data, obtained in FASTQ format, were examined for adapter contamination, low-quality bases, and uncertain nucleotides. Paired reads were discarded if either read contained adapter contamination, more than 10% of bases were uncertain, or if the proportion of low-quality bases exceeded acceptable thresholds (Phred quality score < 5). These quality control steps ensured that only reliable data were used for downstream analysis.

RNA-seq analysis was conducted based on R package DESeq2 (v1.30.0) and clusterProfler (v4.9.0). Gene set enrichment analysis (GSEA) was used to explore the enrichment of KEGG pathways and biological processes within differential expression results. Genes with $|\log_2 \text{FoldChange}| > 1$ and adjusted $P$ value < 0.05), as differentially expressed genes (DEGs), were kept for further analysis. Enrichment analysis was used to assess the enrichment levels of target genes in KEGG pathways and biological processes. Immune cell fractions were inferred by CIBERSORT (Newman et al., 2015) into 22 sub classes with default parameters.

**Statistical analysis**

This study utilized Poisson binomial, Pearson correlation, and hypergeometric distribution statistical models. Fisher's exact test, with Benjamini-Hochberg correction for multiple testing, was the primary method applied for statistical analysis.

## Results

### Integrated compound identification and target validation reveals CB's multi-component synergy

Previous study of CB revealed a total of 49 chemical components, classified into nine distinct groups according to their chemical structures (see Supplementary File). The compound groups ordered by descending proportion were as follows: flavonoids (29%), terpenoids (18%), lignans (16%), coumarins (10%), phenylethanoid glycosides (aka. PeGs) (8%), phenolic acid (6%), volatile oil (8%), polyacetylene (4%), and amino acid (2%) (Figure 1A). To elucidate the distribution of compounds across various herbs, we employed an UpSet plot (Figure 1B) to visualize the number of compounds identified within each herb and their common presence across different herbs. The results demonstrated that the majority of these compounds are derived from *Scleromitrion diffusum* (BHSSC), *Eleutherococcus senticosus* (CWJ), *Poria* (FL), *Atractylodes macrocephala Koidz* (BZ), *Codonopsis radix* (DS), *Reynoutria japonica* (HZ), *Lycium chinense* (GQZ), *Cuscuta* (TSZ), *Schisandra chinensis* (WWZ), *Ligustri lucidi* fructus (NZZ). Notably, certain chemical components, such as Quercetin, are found across multiple herbs exemplifying shared components. This compound, for instance, is present in five herbs used in the CB highlighting its widespread relevance in these herbs.

Based on the components in CB, the DrugCIPHER algorithm was employed to predict the target profile of each identified compound. A comparative analysis between the predicted target profile and those documented in the literature revealed an overlap of over 80%, providing strong validation for the algorithm's predictive accuracy (Figure 1C). Specifically, for syringin, 82% of the predicted targets were either directly supported or indirectly connected through documented protein interactions and signaling pathways. The remaining 18% were linked within a complex biomolecular network, further corroborating the reliability and robustness of the DrugCIPHER predictions (Figure 1D).

**Transcriptomic analysis elucidates immune dysfunction in low-dose radiation-induced leukopenia**

The spleen, as the largest secondary lymphoid organ playing a central role in immune regulation (Lewis et al., 2019) and hematopoiesis (Short et al., 2019), suffers radiation-induced microenvironmental damage that critically contributes to leukopenia development by disrupting both immune regulation and blood cell regeneration (Ackerman et al., 1954; Hamed et al., 2023; Wang et al., 2023). To investigate the mechanisms of low-dose radiation-induced leukopenia, we obtained and analyzed the RNA-seq data of mice spleen before and after radiation exposure. Differential expression analysis identified 155 genes that were significantly downregulated and 108 genes that were upregulated in the irradiated group compared to controls (Figure 2A). GSEA further suggested that radiation exposure disrupts key pathways involved in immune regulation and hematopoiesis (Figure 2B). Notably, immune-related pathways, such as B cell-mediated immunity and B cell activation, were significantly enriched, emphasizing the impact of low-dose radiation on splenic immune function.

To further investigate cellular changes underlying these transcriptomic alterations, immune cell deconvolution of the RNA-seq data was performed to assess proportional shifts among 22 immune cell subtypes within the spleen (Figure 2C). Radiation exposure resulted in significant alterations in the proportions of specific immune cell populations, including a marked reduction in monocytes, macrophages, and naïve $CD4^+$ T cells (Figure 2D). These findings suggest that low-dose radiation disrupts immune cell composition and homeostasis in the spleen, impairing its immune regulatory functions. Collectively, these results provide mechanistic insights into the development of leukopenia and the associated immune dysfunction following radiation exposure (Cytlak et al., 2022; Pandey et al., 2005; Seifert et al., 2016).

**Network target analysis elucidates CB's multi-pathway mechanism governing regulation of leukopenia through herb-target-function triad**

Integrating the GSEA results with enrichment analysis of CB's holistic target, we identified that CB modulates leukopenia through 22 potential pathways encompassed within three classes: cellular process, immune response, and signal transduction (Figure 3A). Within cellular process, CB impacts the cellular activities of B cells, leukocytes, and myeloid cells, notably influencing myeloid cell homeostasis to maintain bone marrow cell stability and promote their growth and differentiation, thereby enhancing leukocyte production. In terms of immune response, pathways such as the immune response-activating cell surface receptor signaling pathway regulate immune responses, augmenting the body's immune function and promoting the activation and proliferation of immune cells. The holistic targets were also enriched in the leukocyte-mediated immunity pathway, indicating the potential of the formulation to enhance the functionality of immune cells such as neutrophils and monocytes, thereby improving the efficacy of the immune system. For signal transduction, CB modulates immune responses and validation processes through pathways such as the interferon-γ mediated signaling pathway and cytokine-mediated signaling pathways. These results indicated that CB demonstrates a multifaceted approach in modulating leukopenia through various pathways, highlighting its potential as a comprehensive therapeutic agent for this condition.

Further analysis of the compounds in CB according to compound types and their respective herbal sources revealed significant intervention roles of coumarin compounds in signal transduction and immune response modules, while terpenes, flavonoids, and volatile oil compounds showed significant activity in the cellular process module (Figure 3B). The herbs in CB can be grouped into four categories based on their roles in disease intervention and their relationships with other herbs in the formula: sovereign (JUN), minister (CHEN) and assistant (ZUO/SHI). In CB, the sovereign herbs are BHSSC and CWJ, while the minister herbs include FL, BZ, DS, GQZ, TSZ, and NZZ, with HZ and WWZ classified as assistant herbs. Enrichment analysis indicated that DS, as a minister herb, demonstrated significant effects across all three modules, thereby assisting the sovereign herbs in enhancing the therapeutic efficacy of CB. Notably, the sovereign herb CWJ, the minister herb NZZ, and the assistant and guide herb WWZ exhibited pronounced activity on the cytokine-mediated signaling pathway (Figure 3B). These three herbs, categorized in TCM as Qi tonics, are known for their ability to promote immune function (Graczyk et al., 2022; Kopustinskiene and Bernatoniene, 2021), as well as their antioxidant and anti-aging properties (Mocan et al., 2014). Building upon the multilayer module network constructed, we identified key targets of CB herbs in the treatment of leukopenia, including TGFB1, C1R, CD4, and HMGB1, which regulate leukopenia through signal transduction, cellular process, and immune response (Figure 4).

**TCM syndromes-guided discovery indicated the multi-target properties of potential bioactive compounds group derived from CB**

Focusing on the TCM syndromes of leukopenia, we screened and identified five compounds significantly intervening in four TCM symptoms, including FD (Lai et al., 2021), WSK (Sungaran et al., 1997), IB and CDFS (Afari and Buchwald, 2003). Based on the established rules for evaluating the intervention of compounds on TCM syndromes, we collected CTD genes associated with these four representative TCM syndromes of leukopenia (see Supplementary File). It was showed that the PeGs-class compounds, including 3-hydroxytyrosol 3-O-glucoside and salidroside, present in the ministerial herb NZZ, along with the terpenoids-class compound syringin, found in the sovereign herb CWJ, exhibit potential intervention effects across all four TCM syndromes of leukopenia (Chen and Fang, 2019). Meanwhile, the lignans-class compound schisandrin, contained in WWZ and the flavonoids-class compound kaempferol-3-glucorhamnoside, found in BHSSC, not only serve as quality control markers for these herbal medicines but also significantly intervene in two key TCM syndromes of leukopenia (IB and WSK, Figure 5A).

Consequently, we identified these five compounds as the bioactive compounds group (BCG) of CB, calculated their collective target profile, and analyzed the differences between the BCG (CB_chemical) and the complete prescription of the capsule (CB_formula). Enrichment analysis revealed seven overlapping pathways shared by CB_chemical and CB_formula, encompassing modules related to immunity, signaling, and cellular processes. These pathways include response to interferon-gamma, activation of immune response, and phagocytosis (Figure 5B).

Moreover, the effects of CB_formula were more significant compared to CB_chemical, which indirectly highlights the synergistic effects of TCM formulas. Furthermore, the BCG was enriched in GO terms associated with leukocytes and lymphocytes, indicating potential intervention effects of the drug combination on cellular processes involving white blood cells and related immune cells (Figure 5D).

We further input the CTD genes associated with the four TCM syndromes of leukopenia into the STRING database to construct a PPI network and subsequently established a multi-level intervention network linking the drug combinations to the syndromes (Figure 5C). In this network, connections between the compound-level modules and syndrome-level modules indicate that the compounds within these modules have potential intervention effects on the corresponding TCM syndromes. The representative compounds from the sovereign herbs and ministerial herbs exhibit potential intervention effects across all four syndromes, while the representative compounds from the adjuvant and guiding herbs demonstrate more pronounced effects specifically on the blood deficiency syndrome.

**CB's bioactive compounds group protected lymphocytes from depletion under irradiation**

To investigate the radioprotective effects of CB multi-component compatibility on immune damage, mice were subjected to 0.1 Gy of ionizing radiation for five consecutive days, with drug administration via gavage immediately following each irradiation session. On the sixth day post-irradiation, immune-related parameters were analyzed in eight groups of mice. These groups included non-irradiated controls (Ctrl), irradiated mice administered normal saline via gavage (IR), and irradiated mice treated with CB multi-component compatibility via gavage (IR+A, B, C or D, Table 1). Each group comprised ten mice. Continuous exposure to radiation resulted in a significant decrease in peripheral blood leukocyte counts in the IR group, affecting lymphocytes, myeloid cells, and megakaryocytes. Notably, treatment with group D demonstrated a protective effect on lymphocyte populations (Figure 6A).

The percentage of T cells among nucleated cells in the spleen exceeds 40%, prompting us to focus specifically on the T cell population. Additionally, we conducted a lineage-level analysis of the thymus to determine whether the T cells present in the spleen originated from this site. Our analysis revealed that group D significantly increased the proportion of T cells within white blood cells in the spleen (Figure 6B-D). Given that the thymus serves as the primary organ for T-cell differentiation, development, and maturation (Shichkin and Antica, 2022), we also tested the variations in T-cell subsets within the thymus across different groups. In the thymus, compared with the IR group, the proportion of progenitor cells was higher in group D than in the drug treatment group; however, the proportion of single positive T cells did not increase (Figure 6E). This suggests that the number of T cells migrating from the thymus to the spleen would not be augmented. The observed increase in the proportion of T cells in the spleen may instead be attributed to enhanced proliferation of naïve T cells. These findings indicate that the protective effect of group D on T cells likely occurs in situ within the spleen.

**In-depth RNA-seq analysis uncovers potential mechanism underlying the intervention of CB's bioactive compounds group in leukopenia**

To investigate the mechanisms underlying the radioprotective effects of the bioactive compounds in CB, we analyzed gene expression profiles in the spleens of mice from the control group (CT), the irradiated group (IR), and the irradiated group treated with CB's BCG (CB). Analyzing the transcriptomic differences between the CB group and IR group, we observed that 161 genes were upregulated and 132 genes downregulated (Figure 7A). The differential gene enrichment analysis confirmed significant effects on 17 of the 22 pathways involved in the full formula intervention of leukopenia (Figure 7B). These pathways impact cellular processes in leukocytes such as lymphocytes and B cells, regulate leukocyte counts, and influence immune-response activating signal transduction pathways such as the B cell signaling pathway, thus enhancing the body's immune function and subsequently elevating leukocyte levels.

Analysis of the differential genes in the CT vs IR and IR vs CB groups revealed that CB's BCG intervention reversed the expression of 16 genes from the model group changes (Figure 7C), and these genes, along with the predicted targets of CB's BCG, were part of the same biological network (Figure 7F). Central to this network, besides immune response cytokines like HLA-B and HLA-A, HSPAIA, MNDA and BST2 are also key roles regulating this network. MNDA is a nuclear protein predominantly expressed in myeloid cells like monocytes and neutrophils (Bottardi et al., 2024; Manohar et al., 2020), involved in immune response and DNA-binding signal transduction (Fotouhi-Ardakani et al., 2010; Meng et al., 2024; Milot et al., 2012). HSPA1A is a heat shock protein involving in protein folding, cell protection, and stress responses (Tóth et al., 2015), regulating apoptosis, cell proliferation, and immune signal transduction. By protecting cells from oxidative stress or cytotoxicity, HSPA1A plays a significant role in maintaining homeostasis and supporting white blood cell proliferation, especially under stress conditions.

Building on insights from flow cytometric analysis, we conducted KEGG pathway enrichment analysis of DEGs and identified four pathways that were simultaneously enriched in both the overall target spectrum of the compound combination and the DEGs, including ovarian steroidogenesis, longevity regulating pathway, chemokine signaling pathway, and cellular senescence (Figure 7D). Furthermore, GO enrichment analysis revealed four significantly enriched terms related to leukocyte proliferation, differentiation, and activation: regulation of leukocyte proliferation, positive regulation of lymphocyte activation, myeloid leukocyte migration, and cytokine-mediated signaling pathway (Figure 7E). Notably, key genes such as MNDA, HSPA1A, HLA-A, HLA-G, and BST2, which are central and densely connected within the biological network, were also involved in these pathways (Figure 7F). These findings suggest that CB's BCG exerts its effects through multiple pathways and gene networks, facilitating leukocyte proliferation and immune recovery.

**Discussion**

Ionizing radiation carries a significant risk of causing leukemia. It has been shown that there is a positive dose-response relationship between long-term exposure to low-dose radiation environments and subsequent death due to leukemia (Leuraud et al., 2015). It is critical to develop therapeutic agents that are effective against leukopenia in exposing to low-dose radiation. However, current commonly used therapeutic options, such as granulocyte macrophage-colony stimulating factor (GM-CSF) and granulocyte-colony stimulating factor (G-CSF), are limited and often ineffective, with significant adverse effects (Chen et al., 2023; Dai et al., 2010).

In clinical settings, the patients commonly manifests with symptoms such as fatigue, dizziness, palpitations, insomnia, poor appetite, low-grade fever, sore throat, and tongue ulcers (GOODMAN et al., 2011). These symptoms, when analyzed through the TCM framework, align with the syndromes of blood deficiency, fatigue syndrome and warm disease. CB, which is known to nourish the kidney (Kinsey and Okusa, 2012) and spleen (Bronte and Pittet, 2013) while enriching and replenishing the blood, has effects that match to these symptom patterns. Studies have demonstrated that CB effectively promote the recovery of peripheral blood counts and hematopoietic colonies after low-dose radiation exposure, significantly increasing WBC levels (Hu et al., 2024).

Through network pharmacology analysis combined with transcriptomic studies(Pang et al., 2021; Wang et al., 2024a), we identified that CB potentially intervene in leukopenia through 22 pathways related to cellular process, immune response, and signal transduction. WBCs are classified as lymphocytes, basophils, neutrophils, eosinophils and monocytes (Craddock Jr et al., 1960). It has been proposed that there is a correlation between leukopenia and changes in the immune milieu, with WBC counts being affected when changes in regulatory T-cell values and IL-7 are produced (Correa-Rocha et al., 2012). Further analysis of the impact of CB components on TCM syndromes such as blood deficiency, fatigue syndrome, warm disease, and spleen-kidney weakness led to the identification of five key compounds as a combination with a significant effect on syndrome modulation. These compounds include kaempferol-3-glucorhamnoside, syringin, schisandrin, 3-hydroxytyrosol-3-O-glucoside and salidroside are all quality controlled compounds of herbs contained in CB. Within them, syringin (Cho et al., 2001; Qian et al., 2024; Us et al., 2015; Zhang et al., 2020) and schisandrin (Fu et al., 2022; Nasser et al., 2020; Wang et al., 2024b) have been found by researchers with medical value in terms of antioxidants and regulating the body’s immunity. Further network target analysis also indicated that this combination notably influences immune signaling pathways and leukocyte- and lymphocyte-related cellular processes.

We determined the effective dosage of this compound combination and validated its efficacy through flow cytometric analysis, confirming its role in increasing peripheral WBC counts and promoting splenic immune cell proliferation in mice. Additionally, RNA-seq analysis revealed that these five key compounds may exert their effects by targeting PP1A, RB, CDK4/6, CDK2, and CDK1, thereby indirectly modulating key molecules such as MNDA, BST2, and HSPA1A. HSPA1A, BST2 and MNDA have high degree in the PPI network, moreover are central genes in functions of controlling the leukocyte cell cycle and affecting its proliferation apoptosis and immunoregulation. These molecules are involved in immune regulation and cellular proliferation, supporting the mitigation of radiation-induced leukopenia through immune and hematopoietic recovery.

Furthermore, we established the association of compounds of CB and potential targets affecting leukopenia by network target analysis (Lai et al., 2020). The positioning of BCG was achieved by combination of TCM syndromes and further verified by ex vivo and in vivo experiments. The result showed that CB's BCG can effectively inhibit radiation-induced lymphocytopenia. We also analyzed the potential intervention mechanisms with RNA-seq results and proposed core targets such as PP1A, RB, MNDA, BST2, and HSPA1A.

Based on a literature review and the gene relationships within cellular senescence, we inferred the regulatory mechanisms by which the potential BCG of CB alleviates leukopenia (Figure 8). Radiation exposure triggers DNA damage through the ATM/ATR-p53 pathways and induces reactive oxygen species (ROS) accumulation (Colin et al., 2014; Li et al., 2024), activating the p38 MAPK pathway to upregulate HSPA1A expression for stress resistance and anti-apoptosis (Shende et al., 2019). Simultaneously, MNDA enhance myelopoiesis via NF-κB signaling and inflammatory cytokine regulation, and BST2 reinforces innate immunity by restricting viral particles and myeloid cell proliferation (Hilligan et al., 2023). CB's BCG synergistically maintains leukocyte homeostasis: HSPA1A preserves cellular homeostasis under oxidative stress, MNDA coordinates differentiation, and BST2 optimizes immune surveillance. Downstream of these targets regulated by CB's BCG, PP1A, RB, CDK4/6, CDK2, and CDK1 are indirectly modulated, leading to influences of cell cycle, promotion of cell proliferation, and resistance to DNA damage. This multi-target defense network regulated by CB integrates immune response and cellular proliferation to combat radiation-induced leukopenia, highlighting the potential of CB in mitigating immune and hematopoietic damage caused by radiation exposure.

Despite the progress made in this study, there are some limitations that should be acknowledged. First, the indication of BCG did not take dosage into account, which was addressed in follow-up experiments to refine the conclusions. Second, the validation of molecular mechanisms was based primarily on transcriptomic data, which provides valuable insights but remains at the transcriptional level. Further experimental studies are needed to confirm the proposed crosstalk at a functional level.

**Conclusion**

This study systematically deciphers the therapeutic mechanism of CB against radiation-induced leukopenia through integrative network pharmacology and experiment validation. From the perspective of the JUN-CHEN-ZUO-SHI classification of TCM, we analyzed the potential mechanisms of the herbs in CB that confer resistance to radiation-induced leukopenia. Through further screening, we identified a BCG comprising five compounds derived from CB. The BCG within CB, dominated by flavonoids and terpenoids, orchestrates HSPA1A, MNDA and BST2 regulatory pathways to counteract radiation damage by modulating cell cycle regulators, resolving DNA damage, and promoting hematopoietic stem cell regeneration. By bridging TCM syndrome patterns with molecular networks, this work establishes a "component-pathway-syndrome" evaluation framework, advancing herbal formula research from empirical application to mechanism-defined therapeutics(张彦琼 and 李梢, 2015). These findings not only validate CB's radioprotective efficacy but also provide a paradigm for modernizing complex traditional formulations through systems biology approaches.

**Data availability**

Raw data is available from the corresponding author upon request. All authors agree to be accountable for all aspects of work ensuring integrity and accuracy.

**Authorship contribution statement**

Dingfan Zhang: Conceptualization, Methodology, Investigation, Formal Analysis, Writing – Original Draft; Congshu Huang: Methodology, Formal Analysis, Data Curation, Writing – Original Draft; Lei Zhou: Validation, Data Curation; Boyang Wang: Writing – Review & Editing; Wei Zhou: Validation, Supervision; Tiantian Xia: Data Curation; Pan Shen: Data Curation, Writing – Review & Editing; Shao Li: Conceptualization, Funding Acquisition, Supervision; Yue Gao: Funding Acquisition, Supervision.

**Acknowledgements**

Not applicable.

**Declaration of competing interest**

The authors declare that they have no known competing financial interests or personal relationships that could have appeared to influence the work reported in this paper.

**Funding**

This work was supported by National Administration of Traditional Chinese Medicine (GZY-KJS-2024-03), the Innovation Team and Talents Cultivation Program of the National Administration of Traditional Chinese Medicine (ZYYCXTD-D-202207).

**Ethics declarations**

**Ethics approval and consent to participate**

All animal experiments were performed according to the Guide for the Care and Use of Laboratory Animals published by the U.S. National Institute of Health (NIH Publication No. 85–23, revised 1996) and were approved by the Bioethics Committee of the Academy of Military Medical Sciences, Academy of Military Sciences (IACUC-DWZX-2025-562).

**Consent for publication**

All authors agreed with the final version and accepted the responsibility to submit for publication.

**Competing interests**

Ding-Fan Zhang, Cong-Shu Huang, Lei Zhou, Bo-Yang Wang, Wei Zhou, Tian-Tian Xia, Pan Shen, Shao Li and Yue Gao declare that there are no potential conflicts of interest that are directly relevant to the content of this article.

# Figures

# Figure 1

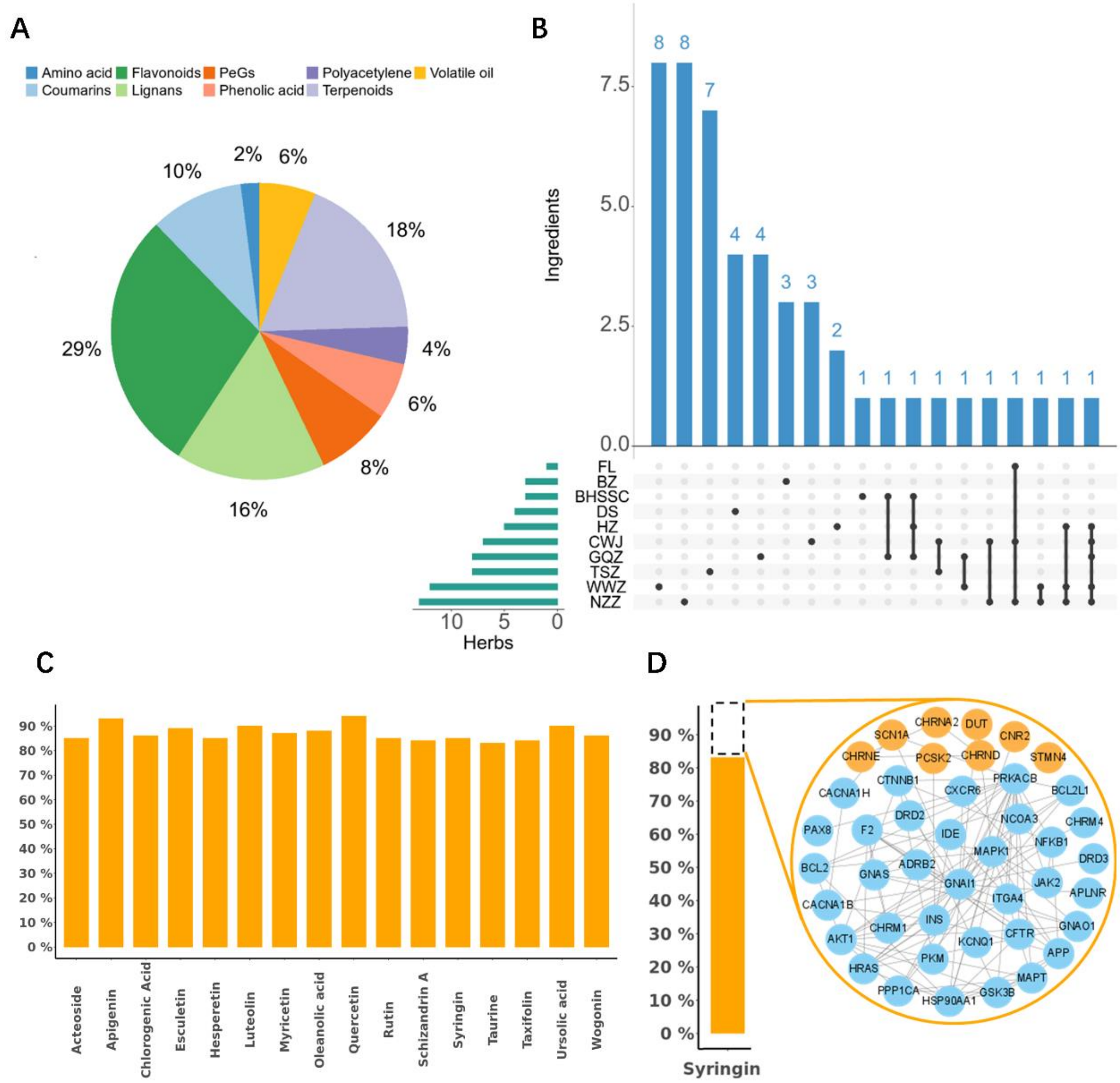

# Figure 2

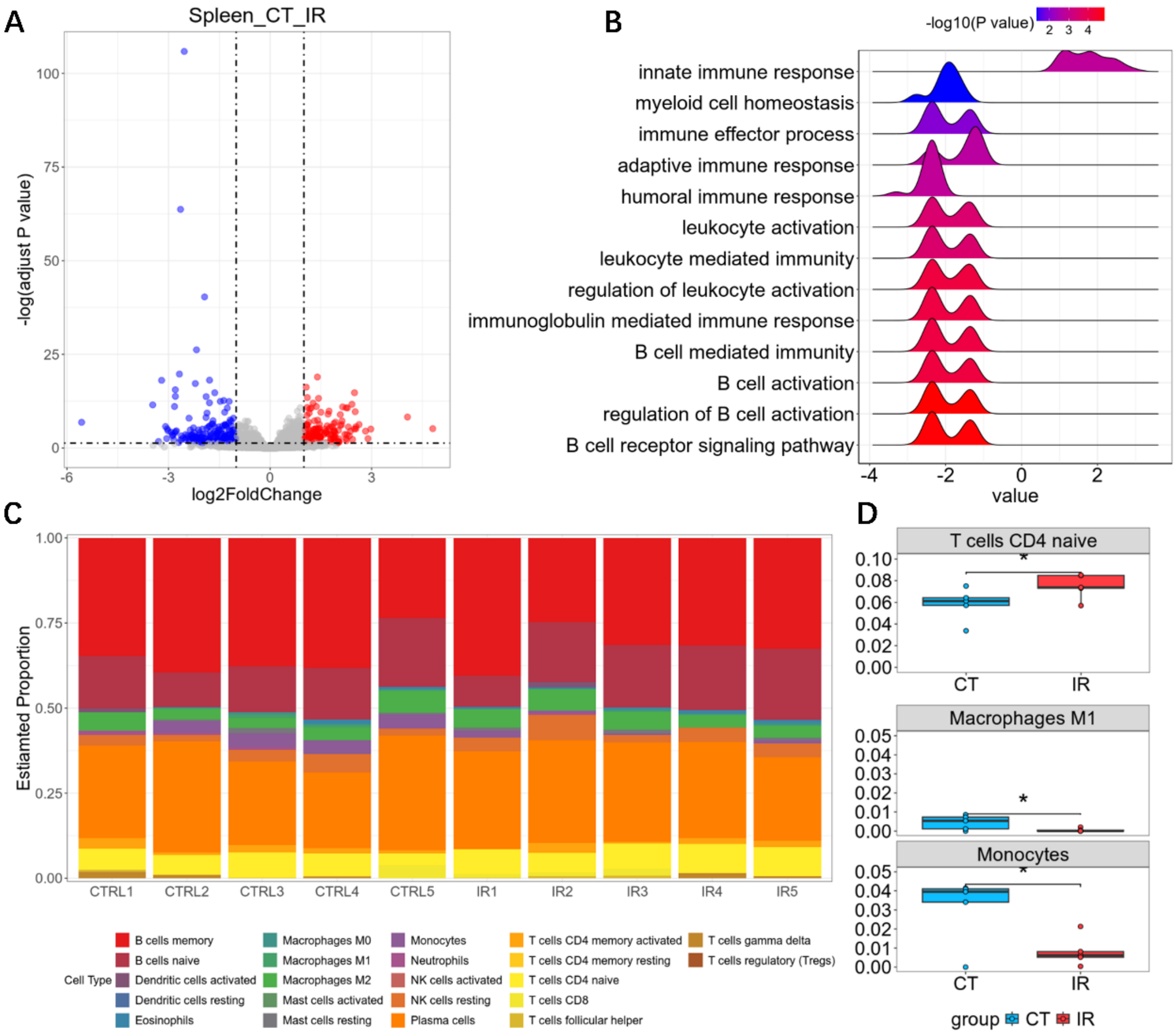
A
Spleen_CT_IR
-log(adjust P value)
log2FoldChange
B
-log10(P value)
innate immune response
myeloid cell homeostasis
immune effector process
adaptive immune response
humoral immune response
leukocyte activation
leukocyte mediated immunity
regulation of leukocyte activation
immunoglobulin mediated immune response
B cell mediated immunity
B cell activation
regulation of B cell activation
B cell receptor signaling pathway
value
C
Estiamted Proportion
CTRL1
CTRL2
CTRL3
CTRL4
CTRL5
IR1
IR2
IR3
IR4
IR5
Cell Type
B cells memory
B cells naive
Dendritic cells activated
Dendritic cells resting
Eosinophils
Macrophages M0
Macrophages M1
Macrophages M2
Mast cells activated
Mast cells resting
Monocytes
Neutrophils
NK cells activated
NK cells resting
Plasma cells
T cells CD4 memory activated
T cells CD4 memory resting
T cells CD4 naive
T cells CD8
T cells follicular helper
T cells gamma delta
T cells regulatory (Tregs)
D
T cells CD4 naive
Macrophages M1
Monocytes
CT
IR
group CT IR

# Figure 3

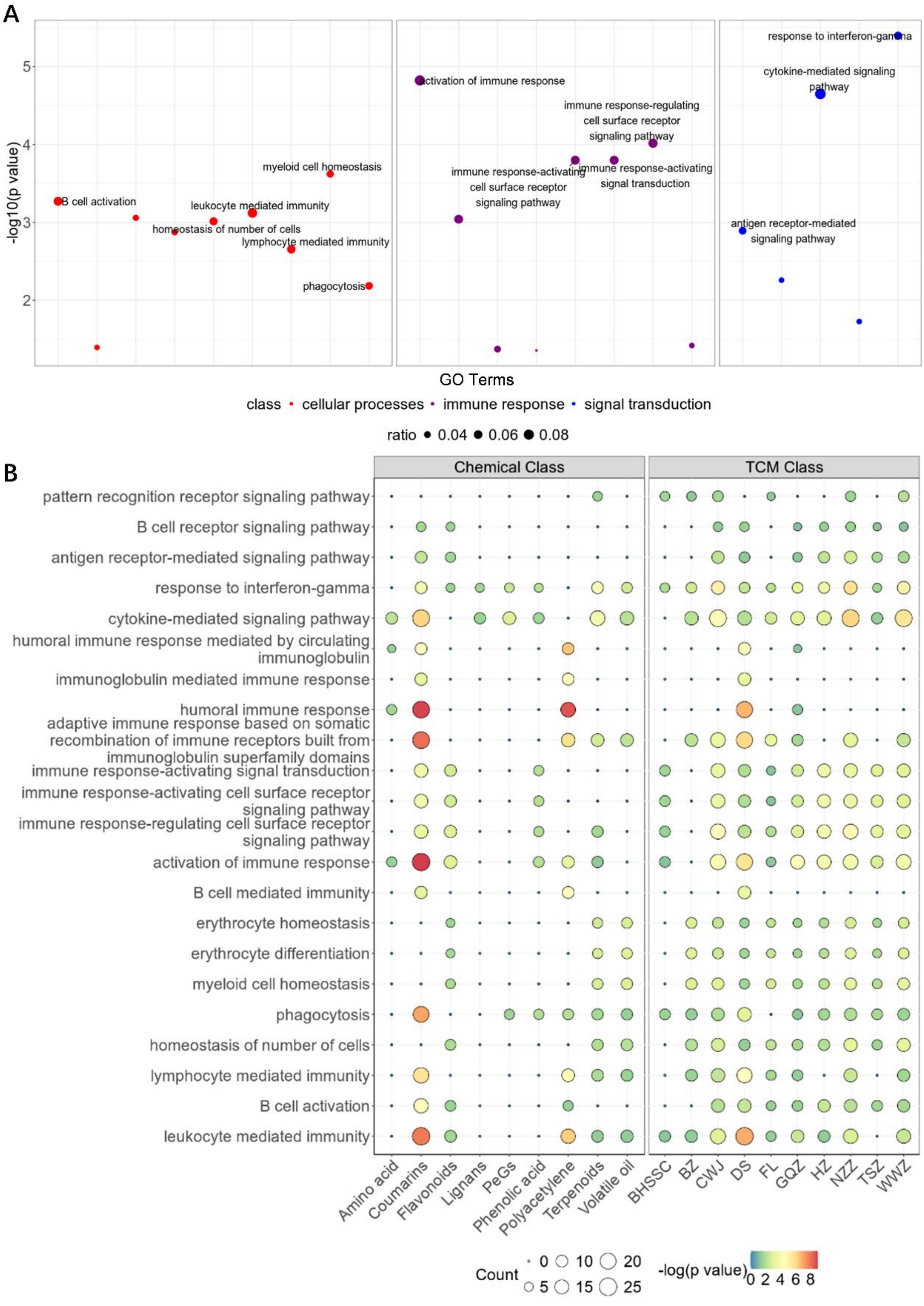
A
response to interferon-gamma
activation of immune response
cytokine-mediated signaling pathway
immune response-regulating cell surface receptor signaling pathway
myeloid cell homeostasis
immune response-activating cell surface receptor signaling pathway
immune response-activating signal transduction
B cell activation
leukocyte mediated immunity
homeostasis of number of cells
lymphocyte mediated immunity
antigen receptor-mediated signaling pathway
phagocytosis
-log10(p value)
GO Terms
class • cellular processes • immune response • signal transduction
ratio • 0.04 • 0.06 • 0.08
B
Chemical Class
TCM Class
pattern recognition receptor signaling pathway
B cell receptor signaling pathway
antigen receptor-mediated signaling pathway
response to interferon-gamma
cytokine-mediated signaling pathway
humoral immune response mediated by circulating immunoglobulin
immunoglobulin mediated immune response
humoral immune response
adaptive immune response based on somatic recombination of immune receptors built from immunoglobulin superfamily domains
immune response-activating signal transduction
immune response-activating cell surface receptor signaling pathway
immune response-regulating cell surface receptor signaling pathway
activation of immune response
B cell mediated immunity
erythrocyte homeostasis
erythrocyte differentiation
myeloid cell homeostasis
phagocytosis
homeostasis of number of cells
lymphocyte mediated immunity
B cell activation
leukocyte mediated immunity
Amino acid
Coumarins
Flavonoids
Lignans
PeGs
Phenolic acid
Polyacetylene
Terpenoids
Volatile oil
BHSSC
BZ
CWJ
DS
FL
GQZ
HZ
NZZ
TSZ
WWZ
Count 0 5 10 15 20 25
-log(p value) 0 2 4 6 8

**Figure 4**

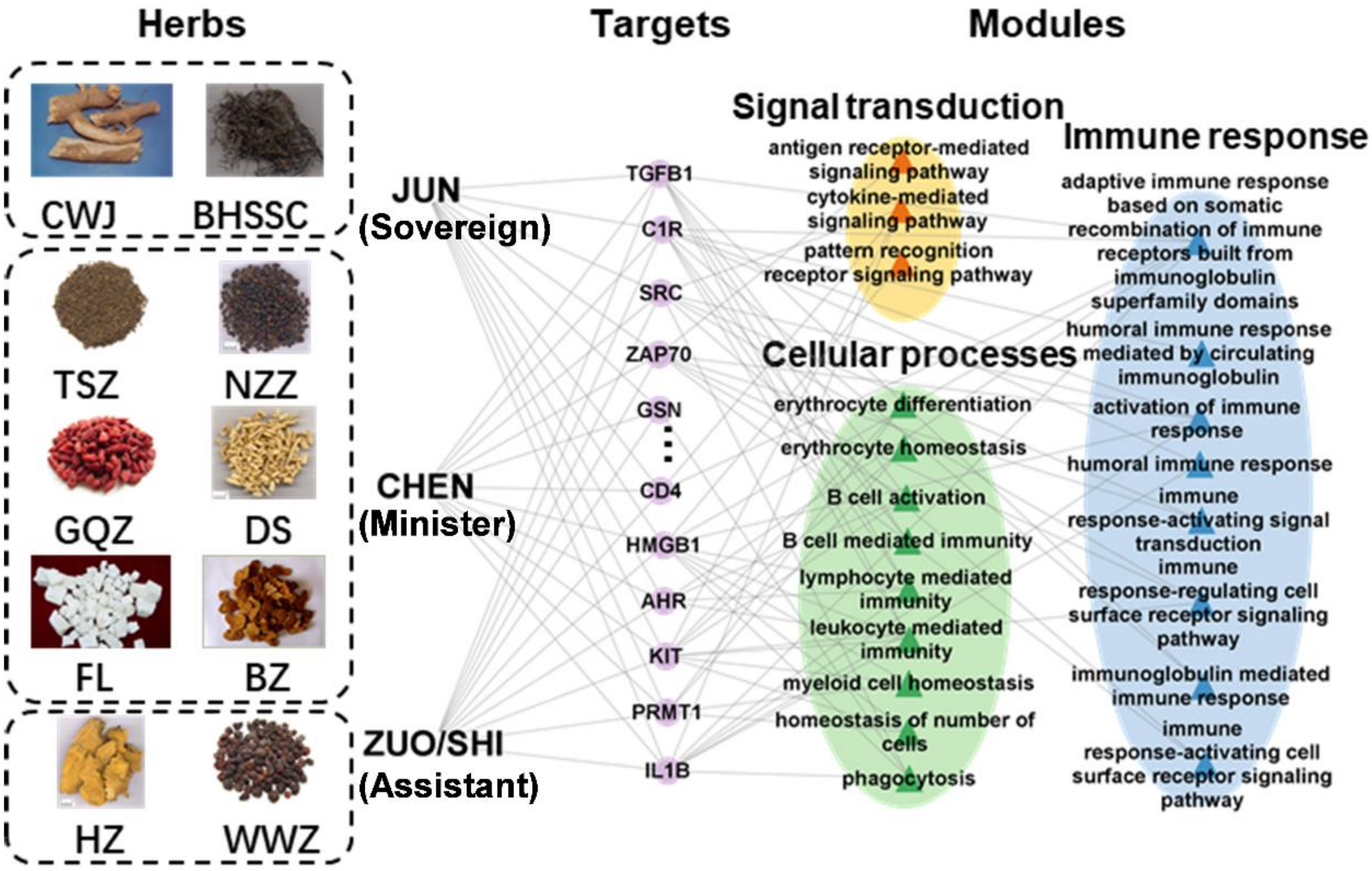

Herbs
Targets
Modules
CWJ
BHSSC
TSZ
NZZ
GQZ
DS
FL
BZ
HZ
WWZ
JUN
(Sovereign)
CHEN
(Minister)
ZUO/SHI
(Assistant)
TGFB1
C1R
SRC
ZAP70
GSN
CD4
HMGB1
AHR
KIT
PRMT1
IL1B
Signal transduction
antigen receptor-mediated signaling pathway
cytokine-mediated signaling pathway
pattern recognition receptor signaling pathway
Cellular processes
erythrocyte differentiation
erythrocyte homeostasis
B cell activation
B cell mediated immunity
lymphocyte mediated immunity
leukocyte mediated immunity
myeloid cell homeostasis
homeostasis of number of cells
phagocytosis
Immune response
adaptive immune response based on somatic recombination of immune receptors built from immunoglobulin superfamily domains
humoral immune response mediated by circulating immunoglobulin
activation of immune response
humoral immune response
immune response-activating signal transduction
immune response-regulating cell surface receptor signaling pathway
immunoglobulin mediated immune response
immune response-activating cell surface receptor signaling pathway

# Figure 5

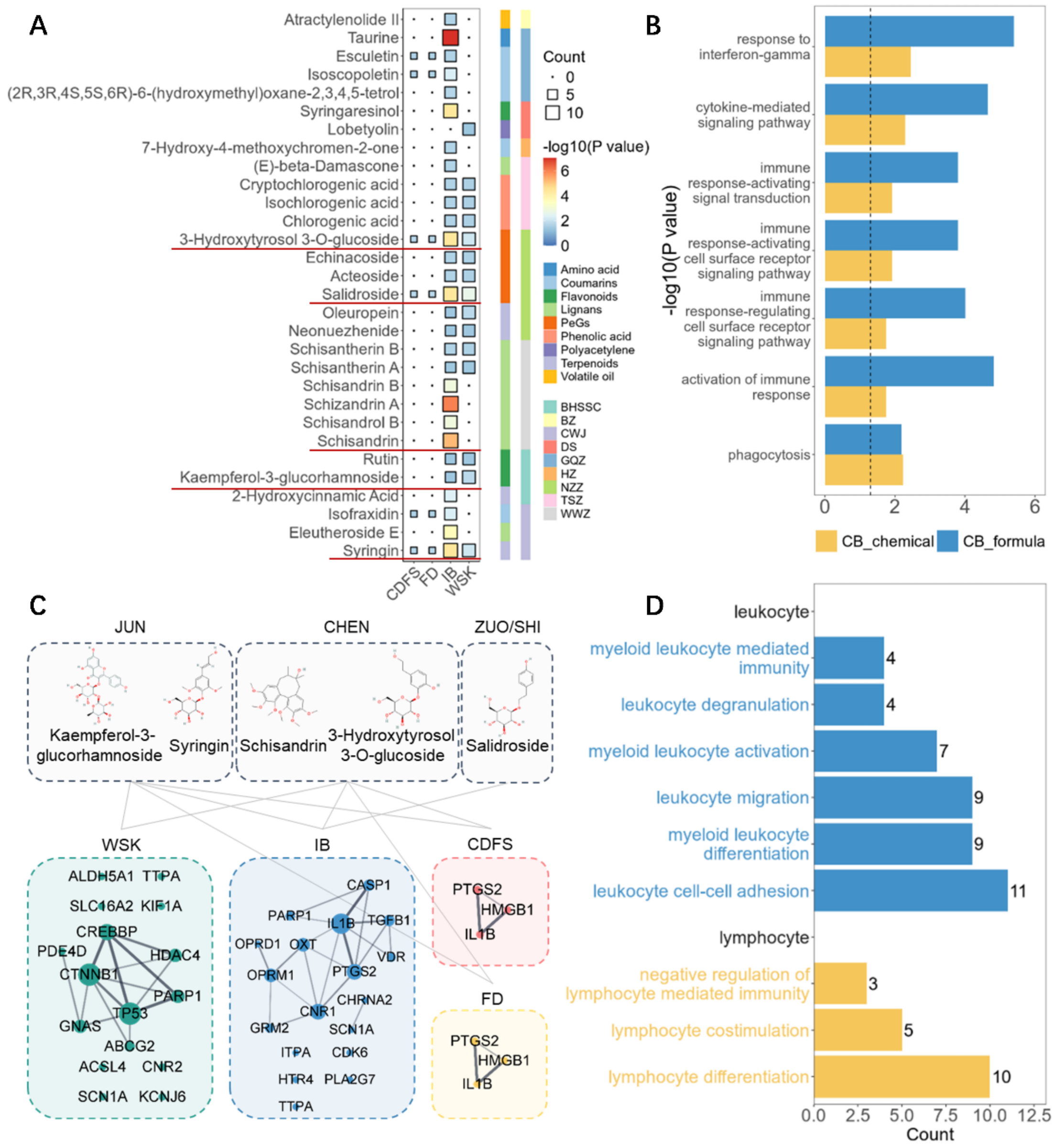

A
B
C
D
Count
-log10(P value)
CB_chemical
CB_formula
JUN
CHEN
ZUO/SHI
WSK
IB
CDFS
FD

**Figure 6**

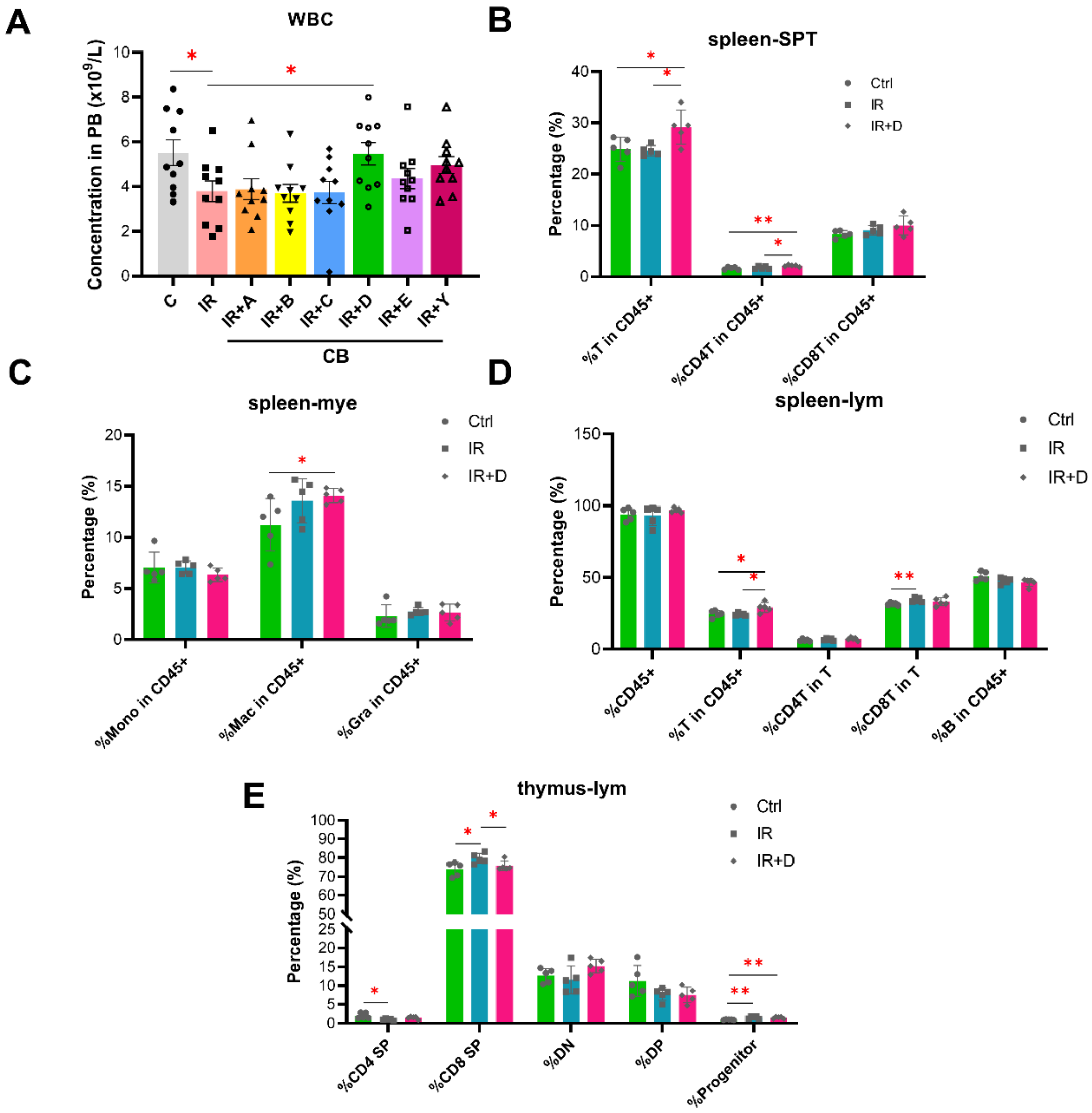
A
WBC
Concentration in PB (x10^9/L)
C
IR
IR+A
IR+B
IR+C
IR+D
IR+E
IR+Y
CB
B
spleen-SPT
Percentage (%)
Ctrl
IR
IR+D
%T in CD45+
%CD4T in CD45+
%CD8T in CD45+
C
spleen-mye
Percentage (%)
%Mono in CD45+
%Mac in CD45+
%Gra in CD45+
D
spleen-lym
Percentage (%)
%CD45+
%T in CD45+
%CD4T in T
%CD8T in T
%B in CD45+
E
thymus-lym
Percentage (%)
%CD4 SP
%CD8 SP
%DN
%DP
%Progenitor

# Figure 7

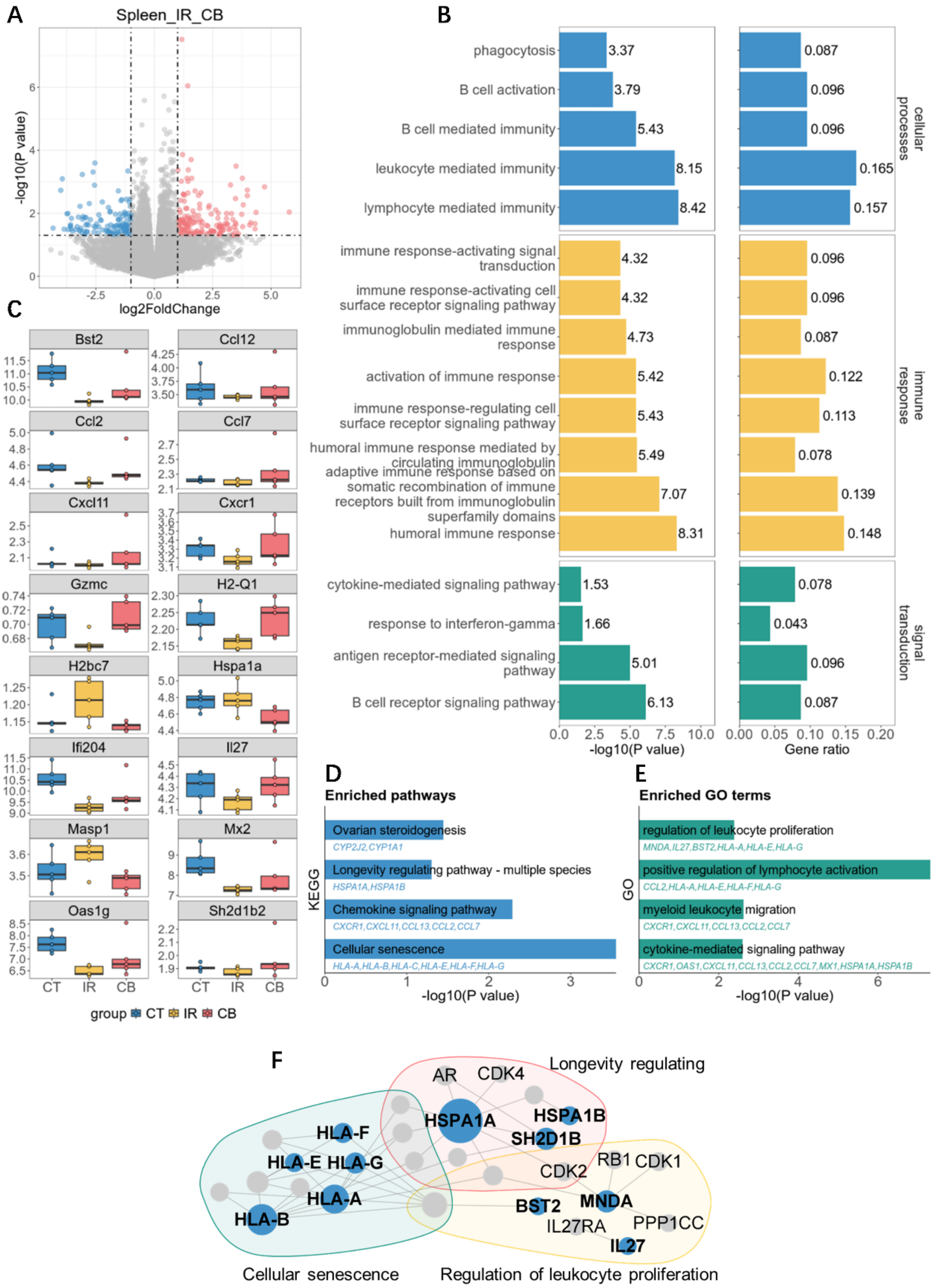

A
Spleen_IR_CB
-log10(P value)
log2FoldChange
B
phagocytosis 3.37 0.087
B cell activation 3.79 0.096
B cell mediated immunity 5.43 0.096
leukocyte mediated immunity 8.15 0.165
lymphocyte mediated immunity 8.42 0.157
cellular processes
immune response-activating signal transduction 4.32 0.096
immune response-activating cell surface receptor signaling pathway 4.32 0.096
immunoglobulin mediated immune response 4.73 0.087
activation of immune response 5.42 0.122
immune response-regulating cell surface receptor signaling pathway 5.43 0.113
humoral immune response mediated by circulating immunoglobulin 5.49 0.078
adaptive immune response based on somatic recombination of immune receptors built from immunoglobulin superfamily domains 7.07 0.139
humoral immune response 8.31 0.148
immune response
cytokine-mediated signaling pathway 1.53 0.078
response to interferon-gamma 1.66 0.043
antigen receptor-mediated signaling pathway 5.01 0.096
B cell receptor signaling pathway 6.13 0.087
signal transduction
-log10(P value)
Gene ratio
C
Bst2
Ccl12
Ccl2
Ccl7
Cxcl11
Cxcr1
Gzmc
H2-Q1
H2bc7
Hspa1a
Ifi204
Il27
Masp1
Mx2
Oas1g
Sh2d1b2
CT IR CB
group CT IR CB
D
Enriched pathways
KEGG
Ovarian steroidogenesis
CYP2J2,CYP1A1
Longevity regulating pathway - multiple species
HSPA1A,HSPA1B
Chemokine signaling pathway
CXCR1,CXCL11,CCL13,CCL2,CCL7
Cellular senescence
HLA-A,HLA-B,HLA-C,HLA-E,HLA-F,HLA-G
-log10(P value)
E
Enriched GO terms
GO
regulation of leukocyte proliferation
MNDA,IL27,BST2,HLA-A,HLA-E,HLA-G
positive regulation of lymphocyte activation
CCL2,HLA-A,HLA-E,HLA-F,HLA-G
myeloid leukocyte migration
CXCR1,CXCL11,CCL13,CCL2,CCL7
cytokine-mediated signaling pathway
CXCR1,OAS1,CXCL11,CCL13,CCL2,CCL7,MX1,HSPA1A,HSPA1B
-log10(P value)
F
Longevity regulating
AR
CDK4
HSPA1A
HSPA1B
SH2D1B
HLA-F
HLA-E
HLA-G
HLA-A
HLA-B
CDK2
RB1
CDK1
BST2
MNDA
IL27RA
PPP1CC
IL27
Cellular senescence
Regulation of leukocyte proliferation

Figure 8

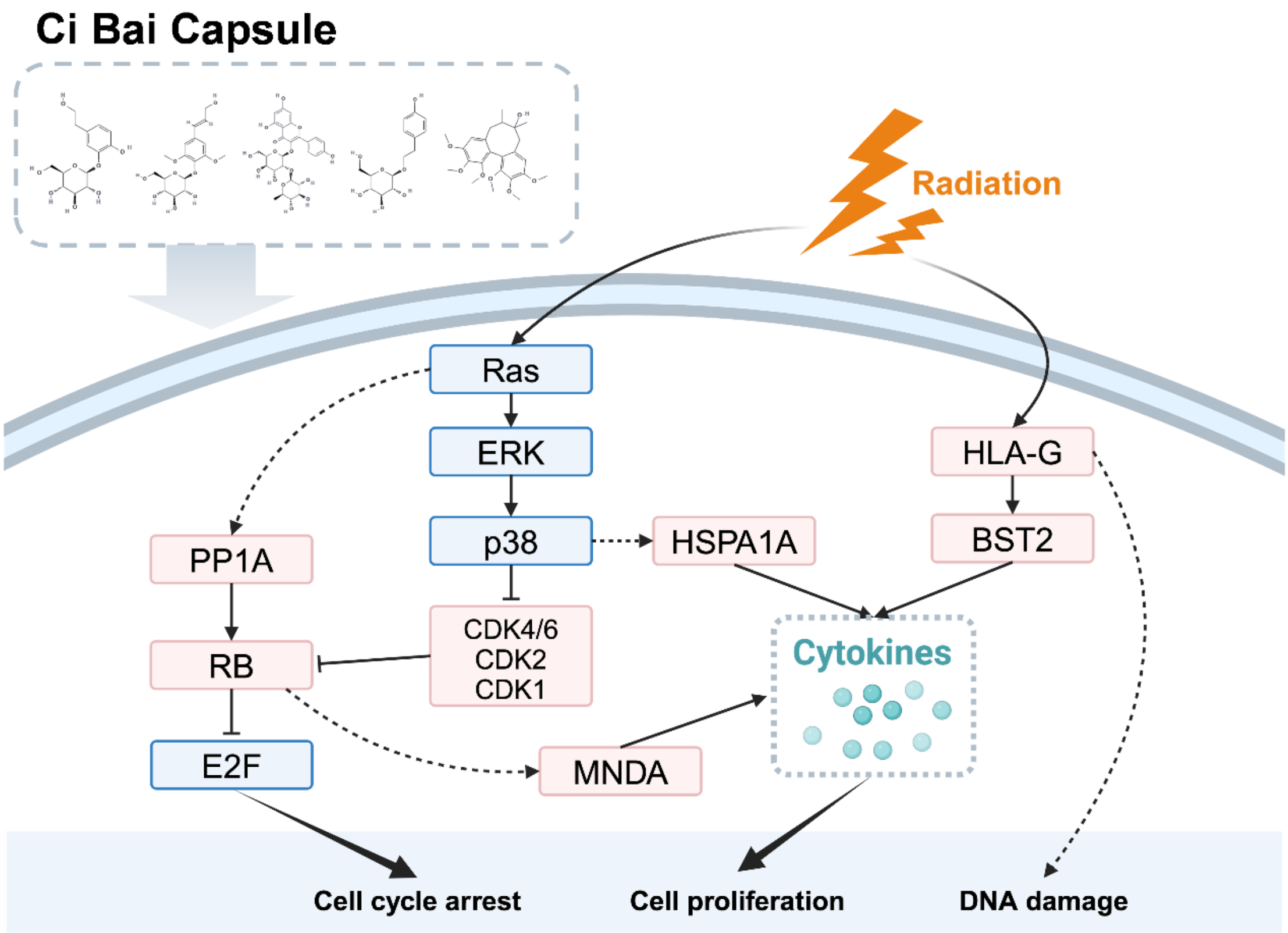
Ci Bai Capsule
Radiation
Ras
ERK
p38
HSPA1A
HLA-G
BST2
PP1A
RB
E2F
CDK4/6
CDK2
CDK1
Cytokines
MNDA
Cell cycle arrest
Cell proliferation
DNA damage

**Figure legends**

**Figure 1. Compound identification, targets prediction and literature validation of predicted targets of compounds in CB.** (A) Pie chart of the chemical classes to which the 49 identified chemical components of CB belonged (B) UpSet plot of the herb classes to which the 49 identified chemical components belonged (C) Target prediction and validation of compounds in CB, displaying literature coverage rate (85–97%) of the main chemical components. (D) The reported and unreported predicted targets of syringin in the same biomolecular network.

**Figure 2. RNA-seq analysis of low-dose radiation-induced leukopenia.** (A) Volcano plots for differential expressed genes before and after radiation in spleen. (B) Enrichment result of the significantly changed biological processes before and after radiation in spleen. (C) Bar plot showing the proportions of immune cells in different samples before and after radiation in spleen. (D) Inferred proportions of naïve T cells, macrophages and monocytes before and after radiation in spleen.

**Figure 3. Network target analysis of CB in modules of leukopenia.** (A) Enrichment result of pathways and biological processes in different modules of CB in the treatment of leukopenia. (B) Enrichment result of pathways and biological processes in different modules of chemical and herb classes in the treatment of leukopenia.

**Figure 4. Multi-layer network representing the potential target of CB in the treatment of leukopenia.** The left boxes listed the herbs in CB classed by JUN, CHEN, ZUO/SHI. The middle dots showed the potential targets of these herbs, and the right triangles indicated functions of these targets.

**Figure 5. Identification and therapeutic effects analysis of potential bioactive compounds group (BCG) of CB.** (A) Heatmap of the potential interventional effects of compounds contained in CB on four syndromes. (B) Differences between potential BCG and the whole formula of CB's intervention in leukopenia. (C) Multi-level intervention network of potential BCG to four syndromes of leukopenia. (D) Significant pathways of potential BCG in leukocytes and lymphocyte.

**Figure 6. CB exhibits a radioprotective effect on splenic lymphocytes.** (A)

Number of peripheral blood leukocytes after CB multi-component compatibility treatment. (B-D) Percentage of T-cell-related populations in the spleen following drug treatment in group D. (E) Percentage of T-cell-related populations in the thymus following drug treatment in group D. DN, double-negative T cell, DP, double-positive T cell.

**Figure 7. RNA-seq analysis of CB's bioactive compounds group (BCG) intervention in leukopenia.** (A) Volcano plots for differential expressed genes before and after the invention of potential BCG of CB (B) Enrichment result of the significantly changed biological processes before and after the invention of potential BCG of CB. (C) The expression of 16 genes significantly changed before and after the invention of potential BCG of CB. (D-E) Enriched pathways and biological terms of significantly changed genes. (F) Biological network between significantly changed genes before and after the invention of potential BCG of CB and network targets of potential BCG of CB.

**Figure 8. Mechanistic pathways of CB's potential bioactive compounds group (BCG) in mitigating radiation-induced leukopenia.** Radiation initiates DNA damage via ATM/ATR-p53 activation and ROS overproduction, triggering p38 MAPK-mediated HSPA1A upregulation for stress adaptation and apoptosis resistance. Concurrently, MNDA promoters NF-κB-driven myelopoiesis while BST2 enhances antiviral immunity through myeloid proliferation. CB's BCG synergistically sustains leukocyte homeostasis by coupling oxidative defense (HSPA1A), differentiation coordination (MNDA), and immune optimization (BST2). CB's BCG amplify these effects by modulating cell cycle regulators (PP1A/RB/CDKs), resolving DNA damage and promoting cell proliferation.

**Table 1. Uniformly designed and screened the experimental prescription administration scheme.**

| Group | Syringin $mg \cdot kg^{-1}$ | Salidroside $mg \cdot kg^{-1}$ | Kaempferol $mg \cdot kg^{-1}$ | Schisandrol A $mg \cdot kg^{-1}$ | Cimidahurinine $mg \cdot kg^{-1}$ |
|---|---|---|---|---|---|
| Uniform 1 (A) | 10 | 25 | 19 | 62 | 41 |
| Uniform 2 (B) | 20 | 55 | 40 | 5 | 33 |
| Uniform 3 (C) | 30 | 85 | 12 | 81 | 25 |
| Uniform 4 (D) | 40 | 10 | 33 | 24 | 17 |
| Uniform 5 (E) | 50 | 40 | 5 | 100 | 9 |
| Uniform 6 (Y) | 60 | 70 | 26 | 43 | 1 |